\documentclass[runningheads,a4paper]{llncs}
\usepackage[width=122mm,left=44mm,paperwidth=210mm,height=193mm,top=52mm,paperheight=299mm]{geometry}

\usepackage{latexsym}
\usepackage{graphicx}
\usepackage{epstopdf}

\usepackage{indentfirst}
\usepackage{booktabs}
\usepackage{array}
\usepackage{float}
\usepackage{stfloats}
\graphicspath{{./figures/}}
\usepackage{threeparttable}
\usepackage{url}
\usepackage{setspace}
 \normalsize

\begin{document}

\title{Dynamic Flow Scheduling Strategy in \\Multihoming Video CDNs}

\author{Ming Ma\inst{1}%
\and Zhi Wang\inst{2} \and Yankai Zhang\inst{1} \and Lifeng Sun\inst{1}}

\institute{Tsinghua National Laboratory for Information Science and Technology\\ Department of Computer Science and Technology, Tsinghua University 
\and
Graduate School at Shenzhen, Tsinghua University\\
\email{\{mm13@mails., wangzhi@sz., zyk12@mails, sunlf@\}tsinghua.edu.cn}}

\maketitle

\begin{abstract}

	Multihoming for a video Content Delivery Network (CDN) allows edge peering servers to deliver video chunks through different Internet Service Providers (ISPs), to achieve an improved quality of service (QoS) for video streaming users. However, since traditional strategies for a multihoming video CDN are simply designed according to static rules, e.g., simply sending traffic via a ISP which is the same as the ISP of client, they fail to dynamically allocate resources among different ISPs over time. In this paper, we perform measurement studies to demonstrate that such static allocation mechanism is inefficient to make full utilization of multiple ISPs' resources. To address this problem, we propose a \emph{dynamic flow scheduling strategy for multihoming video CDN}. The challenge is to find the control parameters that can guide the ISP selection when performing flow scheduling. Using a data-driven approach, we find factors that have a major impact on the performance improvement in the dynamic flow scheduling. We further utilize an information gain approach to generate parameter combinations that can be used to guide the flow scheduling, i.e., to determine the ISP each request should be responded by. Our evaluation results demonstrate that our design effectively performs the flow scheduling. In particular, our design yields near optimal performance in a simulation of real-world multihoming setup.
	

\end{abstract}

\section{Introduction}

Content Delivery Networks (CDNs) have been massively used by video service providers to deliver video chunks using the geo-distributed servers. In such video CDNs, multihoming, that allows one peering server\footnote{CDN edge peering servers are server clusters located at the edge of the network to which end users are connected.} to deliver video chunks through networks of different Internet Service Providers (ISPs) to different users (referred to as \emph{flow scheduling} in this paper), has become a promising way to improve the quality of services (QoS) for data-intensive video delivery, e.g., improving the service availability \cite{liu2007survey}. 

\begin{figure}[t]
\begin{minipage}[t]{\linewidth}
    \centering
     
    \includegraphics [width=0.45\linewidth,natwidth=610,natheight=642]{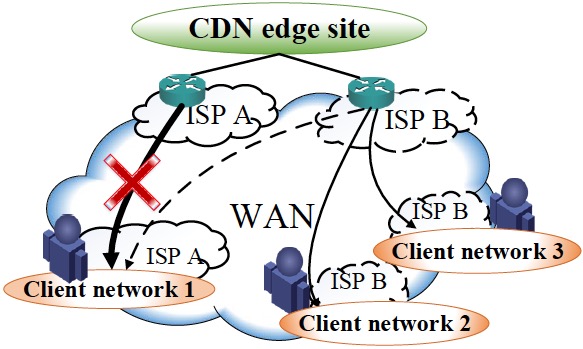} 
  
  \end{minipage}
 \hfill
 
  \caption{A simple case to illustrate the inefficiency of rule-based strategy. }
  \label{fig:example}
  \vspace{-4mm}
\end{figure}

In multihoming CDNs, traditional approaches schedule flows using static rule-based strategies, e.g., a user is scheduled to receive data from the CDN server's outgoing ISP which is the same ISP as the user's ISP (referred to as same ISP-based strategy) \cite{PeeringNetwork}\cite{bindal2006improving}. Such static rule-based strategies fail to make full use of network resources that are dynamically changing over time and across locations \cite{akella2003measurement}\cite{genin2013internet}.
 Fig.~\ref{fig:example} is a simple case that illustrates this inefficiency.
 In this case, a CDN edge site, multihomed to $2$ ISPs, delivers videos with the same ISP-based strategy. Thus even when the path ``CDN network (ISP A) $\rightarrow$ Client network (ISP A)'' has congestions, the peering server still delivers the contents along this congested path ignoring a better choice ``CDN network (ISP B) $\rightarrow$ Client network (ISP A)''. 

Thus it is intriguing to study dynamic strategies for flow scheduling in multihoming video CDNs. In this paper, we carry out measurement and strategy designs to answer the following questions: (a) How much gain can we achieve from the dynamical flow scheduling strategy compared to a static one in multihoming video CDN? (b) Which factors will influence the performance gain? And (c) how can we design strategies that can be used by today's CDNs to achieve such performance gain?

To answer questions (a) and (b), we use both real-world and simulation experiments to verify the advantages of using dynamical strategies to allocate network resources, and investigate the network factors which affect the performance of dynamical flow scheduling, e.g., the hops between the server and clients. For question (c), based on the measurement insights, we select a set of parameters, and these parameters can be used by video CDN providers to actively and dynamically control \emph{which flow will go through which ISP's network from a peering server at a given time}. 

Our contributions are summarized as follows:

$\rhd$  We conduct extensive real-world and simulation experiments to confirm the fact that dynamical flow scheduling significantly outperforms conventional static strategies in multihoming CDNs, i.e., in the controlled simulation experiments, we observe that the throughput of dynamical strategy can even be $2$x higher than the static strategy.
		
$\rhd$  We further identify the parameters that affect the performance gain by observing and comparing network information under all possible flow schedule strategies, and find the most influential parameters are link coverage, overlap and the total bandwidth of bottlenecks.
	
$\rhd$  We design a local selection scheme using a combination of parameters, to determine which flow goes through which ISP's network. Our evaluation verify the effectiveness of our design, yielding near optimal performance (up to $90\%$ of the optimal gain) in most of scenarios under the real-world multihoming simulation.

The rest of this paper is organized as follows. Sec.~\ref{sec:relatedwork} discusses the related works. Sec.~\ref{sec:def} gives some key definitions in this paper. In Sec.~\ref{sec:measure}, we carry out real-world measurements to confirm that dynamic flow scheduling is a must in today's multihoming CDN. In Sec.~\ref{sec:topologyanalysis} we investigate the factors that affect the performance gain of dynamic flow scheduling. In Sec.~\ref{sec:design}, we design a local selection scheme to schedule flows and evaluate our design in Sec.~\ref{sec:evaluation}. We conclude the paper in Sec.~\ref{sec:conclusion}.

\section{Related Works} \label{sec:relatedwork}

 Video content constitutes a dominant fraction of online entertainment traffic today. There are abundant techniques \cite{baochun-tomccap-streaming2013}\cite{cp_cdn} used in the content-distribution systems.

QoS for video chunk delivery has become a key point to improve the quality of user experience in video streaming systems~\cite{liu2012case}. To this end, multihoming has becoming the mainstream solution for CDN providers to improve the service reliability and performance. In this multihoming CDN context, rule-based strategies have been used to schedule flows in a static way due to the charge between ISPs or other commercial polices~\cite{PeeringNetwork}\cite{bindal2006improving}. However, according to Akella et al.'s study~\cite{akella2003measurement}\cite{4444747}, strategical scheduling traffic across the ISPs can improve performance by $25\%$. And Valancius et al.~\cite{valancius2013quantifying} claim that an online service provider can provide tangible improvement ($4.3\%$) over the state of the art by controlling both the selection of CDN sites and the multiple ISPs between the clients and their associated CDN sites.

As multihoming is a promising approach to enhance the performance of content delivery, some existing works explore the problem of dynamic flow scheduling. Akella et al.~\cite{4444747} choose the outgoing ISP with the least round-trip time (RTT), and propose a method to reduce the ping probes in the link-performance monitoring process of the route optimization. Goldenberg et al.~\cite{goldenberg2004optimizing}\cite{dhamdhere2006isp} develop algorithms for optimizing both cost and RTT for multihomed networks.

These existing studies are not suitable for the CDN of data-intensive video streaming, which pays more attention to high throughput. How to design a more practical strategy to schedule flows for a video peering server is still a challenging problem. In this paper, we study parameters that can be used in real-world flow scheduling for multihoming video CDNs.

\section{Key definitions}
\label{sec:def}
Before we present the experiments to motivate the dynamic flow scheduling in multihoming CDN, we define three important terms used in this paper.

$\rhd$ \textbf{Multihoming ISP and client network.} $\mathcal{F}$ denotes the set of ISPs to which the peering server multihomed. $f \in \mathcal{F}$. $\mathcal{C}$ denotes the set of client networks. $c \in \mathcal{C}$. 

$\rhd$ \textbf{Flow schedule strategy}. We define the flow schedule strategy as an {\emph{outgoing ISP vector}. Let $k$ denote a flow scheduling strategy. $k=(f_{1}, f_{2}, ..., f_{\mathcal{C}|})$, where $f_{i} \in \mathcal{F}$ indicates the outgoing ISP selected for the $i^{th}$ client network. To explore the optimal performance gain of flow scheduling, we assume the CDN peering server can enumerate all possible flow strategies. The number of flow scheduling strategies is $|\mathcal{F}|^{|\mathcal{C}|}$. 

$\rhd$ \textbf{Performance gain}. On account of the characteristic of video streaming, we focus on the evaluation of the throughput improvement achieved by dynamic flow scheduling. And the performance gain is defined as the maximal throughput achieved within all of the flow scheduling strategies divide the throughput achieved by a static flow schedule strategy. 

\section{Motivation: Real-world experiments}
\label{sec:measure}

In this section, we carry out real-world experiments to assess the performance gain of dynamic scheduling in multihoming CDN. 
We deploy $2$ peering severs in a Internet Data Center (IDC) site of Cycomm, a commercial IDC provider in China, to simulate a $2$-multihomed CDN server. These $2$ peering servers are connected to the ISP China-Unicom (labeled as ISP A) and the ISP China-Telecom (labeled as ISP B) respectively. We have $2$ client networks, i.e., Tsinghua campus network (labeled as Client Network $1$) and an ADSL network (labeled as Client Network $2$). All of the server and client networks are deployed in Beijing. As illustrated in Fig.~\ref{fig:3-2realtrace}(a), The flow schedule then has the following strategies, $(A, A)$, $(A, B)$, $(B, A)$ and $(B, B)$. For example, $(A, B)$ represents that the server delivers chunks to Client network $1$ using ISP A and delivers chunks to Client network $2$ using ISP B. We carry out experiments for $14$ days. Every $15$ min, all clients concurrently download video chunks whose size varies from several megabytes to tens of megabytes from the peering server using 4 strategies in turns, and we record the download speed of each strategy. 

\begin{figure*}[tbp]
 \begin{minipage}[t]{0.320\linewidth}
    \centering
    \includegraphics [width=\linewidth]{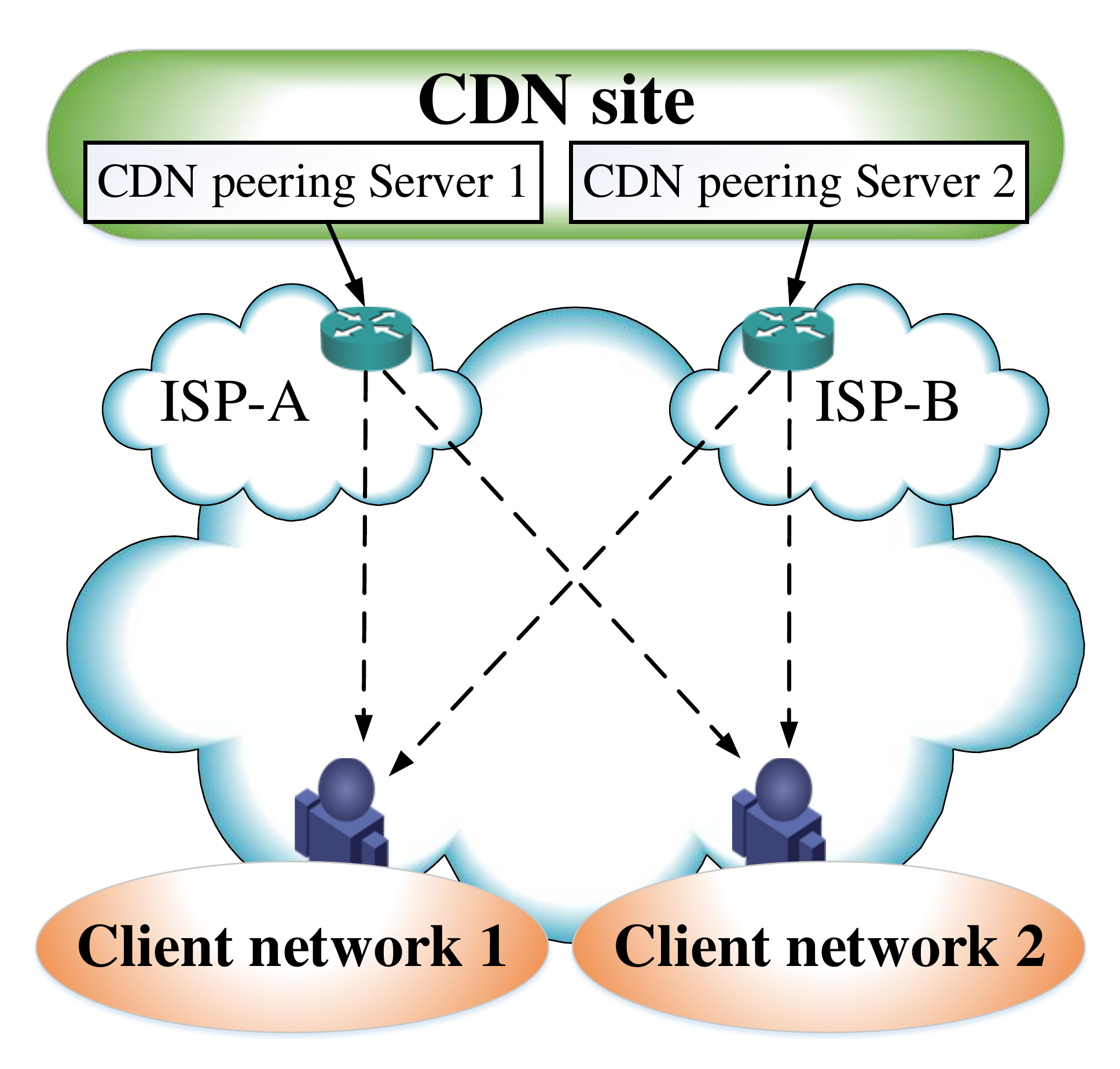} 
    \centerline{\scriptsize (a) Experiment setup}
  \end{minipage}
  \hfill
  \begin{minipage}[t]{0.32\linewidth}
    \centering
    \includegraphics [width=1.1\linewidth]{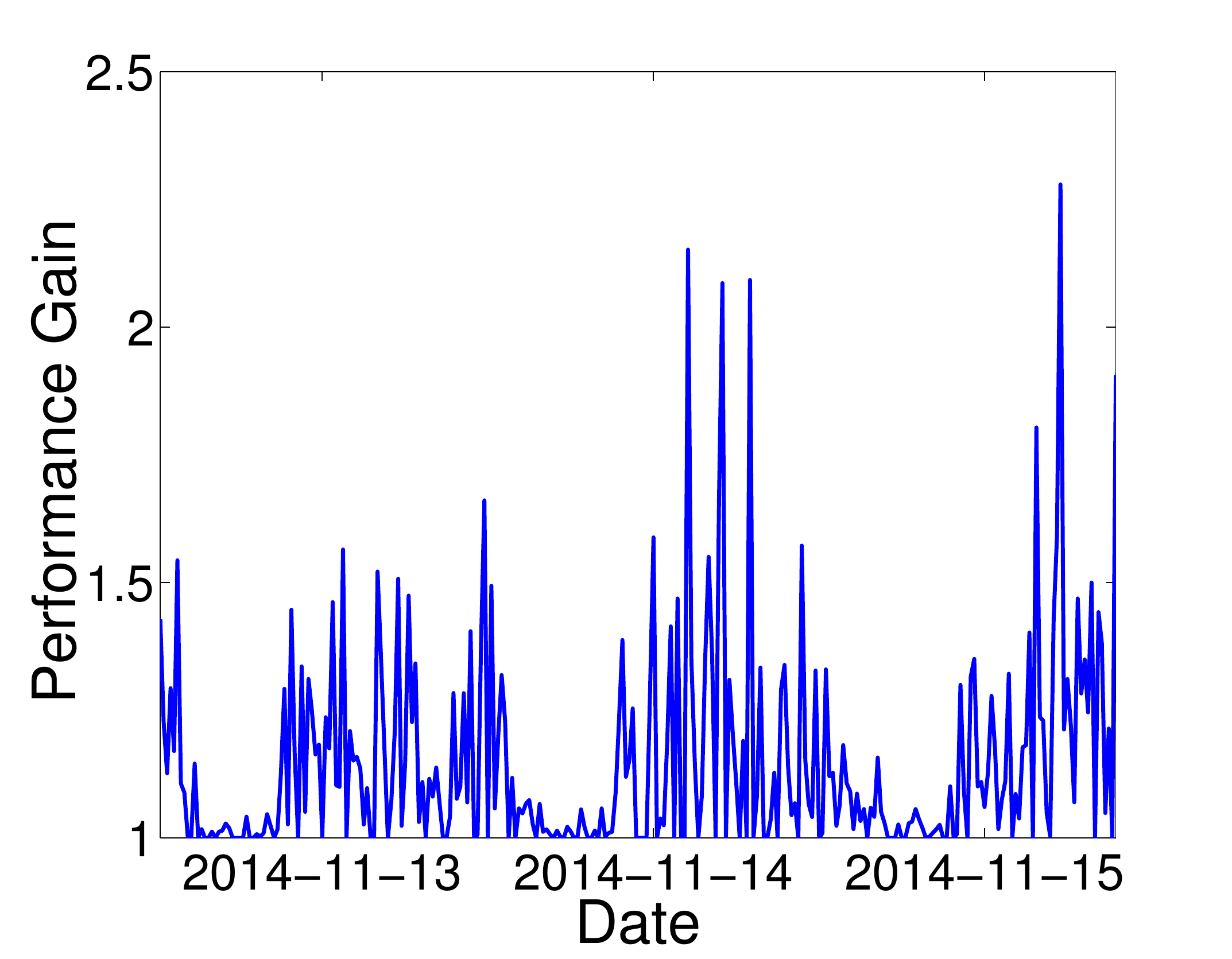}
    \centerline{\scriptsize (b) Throughput Gain}
  \end{minipage}
  \hfill
    \begin{minipage}[t]{0.32\linewidth}
    \centering
    \includegraphics [width=1.15\linewidth]{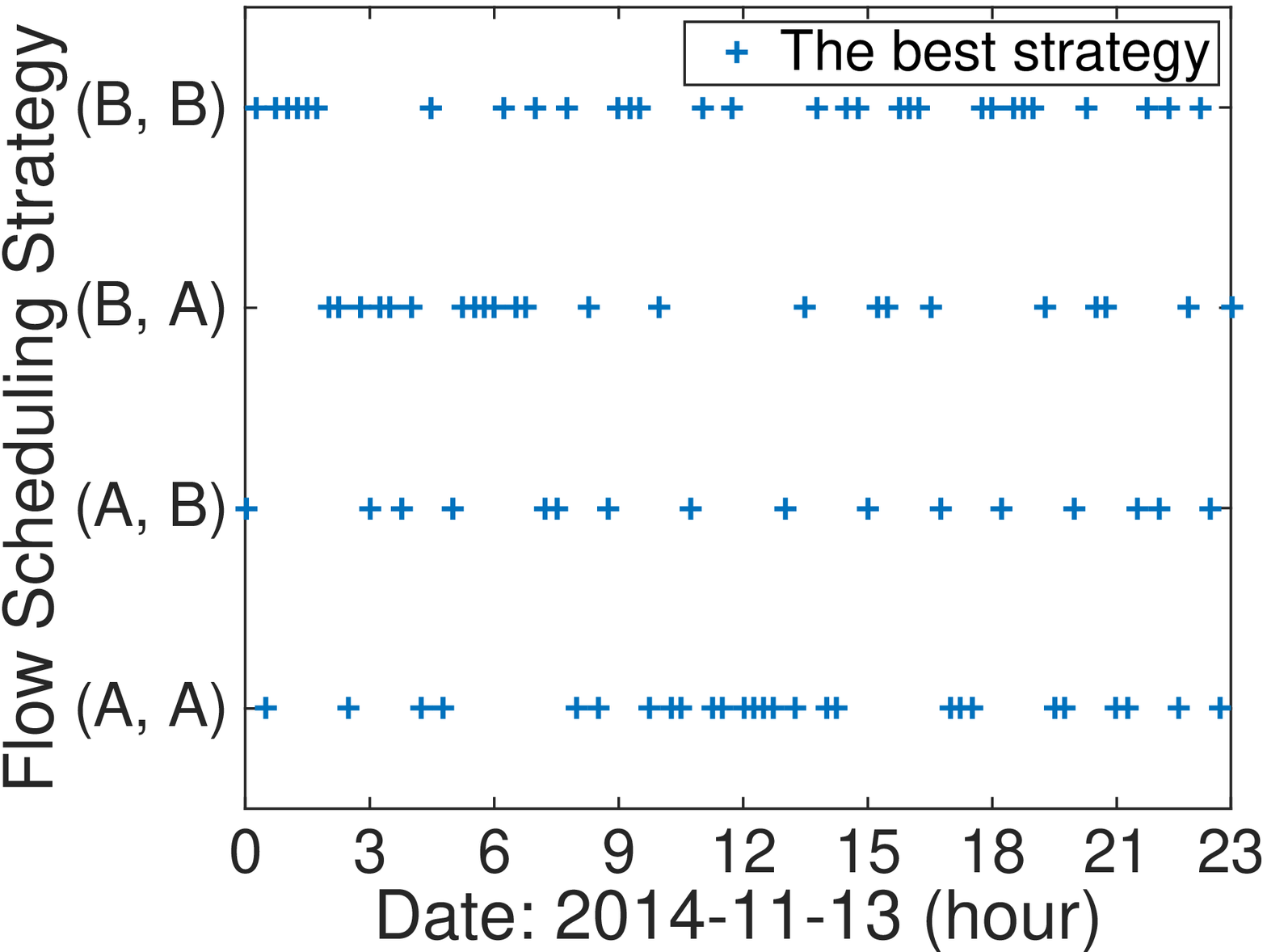}
    \centerline{\scriptsize (c) Optimal Strategy Distribution}
  \end{minipage}
  \hfill
   \vspace{-1mm}
  \caption{Real-world experiment with controlled clients.}

  \label{fig:3-2realtrace}
  \vspace{-4mm}
\end{figure*}

Based on the traces recorded for the two controlled client networks, we plot the performance gain over three days in Fig. \ref{fig:3-2realtrace}(b). The performance gain is that the maximal throughput achieved among all of the four strategies divides the throughput achieved by a static flow schedule strategy (always using $(A, B)$ in our calculation). Our observations are as follows: (1) The best flow schedule strategy in multihoming CDN can significantly improve the throughput to both client networks (up to $2$x); (2) The performance gain is varying over time following a daily pattern, and large gain is achieved during busy hours (daytime). 

Furthermore, we investigate which strategies achieve the largest throughput over time. As illustrated in Fig.~\ref{fig:3-2realtrace}(c), we plot the \emph{best strategies} for each time slot ($15$ min). We observe that the best strategy varies significantly over time and may be different in consecutive time slots, which indicates that in a real-world network condition, multihoming CDN needs to schedule flows in a dynamic manner by adjusting outgoing ISPs.

\section{Performance Gain in Multihoming Video CDN}
\label{sec:topologyanalysis}

 In this section, we analysis the performance gain of flow scheduling in different scenarios. We perform the controlled simulations to study the factors that intrinsically affect the the performance gain in multihoming CDN.

\subsection{Simulation Settings} \label{controlled simulation set}

$\rhd$ \textbf{Network Topology.} Our simulation experiments are based on two types of network topologies: 1) Generated topologies. We use BRITE \cite{medina2001brite}, a topology generation tools, to create different network topologies with the algorithms \emph{Waxman} (a flat topology model) and \emph{Transit-Stub} (referred to as TS, a hierarchical topology model); 2) Real-world topologies. We use four public datasets of network topologies (referred to as Real 1, Real 2, Real 3 and Real 4), from Route Views Project and Caida Website\footnote{http://www.routeviews.org/; http://www.caida.org/} in our experiments. 

$\rhd$ \textbf{Network Link Capacity.} Due to the distribution of the link capacity is hard to know \cite{Yu:2012:TCD:2413176.2413194}, we set the link capacities in the network following the heavy-tailed distribution ranging from $1$ Gbps to $50$ Gbps.

$\rhd$ \textbf{In-network Routing.} Conventional OSPF strategy is used to routing the packets between other nodes in our experiments, and the each link costs is setted as $10^{8}/Bandwidth$, following the default configuration of Cisco\footnote{http://www.cisco.com/c/en/us/support/docs/ip/open-shortest-path-first-ospf/7039-1.html}. 

$\rhd$ \textbf{Multihoming CDN Settings.} $500$ different topologies are generated for the same configuration (i.e., topology structure, client request, etc.) in our experiments. For each topology, we randomly choose one node which has three edges as the $3$-multihoming CDN peering server, i.e., $|\mathcal{F}| = 3$ and $\mathcal{F}=\{A,B,C\}$, and $5$ other nodes as client nodes (a client node represents a client networks in the simulation), i.e., $|\mathcal{C}| = 5$ and we set the static flow strategy $k$ as $(A,A,B,B,C)$ for each topology simulation. 

$\rhd$ \textbf{Throughput Calculation under a Flow Scheduling Strategy.} Given a topology with CDN servers and multiple client networks, we can calculate the server throughput of any flow scheduling strategy by formulating this question as a traffic engineering problem in \cite{altin2013intra}, constrained by the capacities of links in the network and the bandwidth requirement of each client.

Note that we carry out other experiment with different setting and the results exhibit similar patterns, thus we only present the results of the simulation settings above in this paper due to page limitation.

\subsection{What Influences the Performance Gain}

We present the impact of different network topologies on the performance gain in Tab.~\ref{tab:topology}. The client demands are set unconstrained (from $0$ to $\infty$) to learn the maximum throughput we can get from the network. In this table, we denote the number of nodes as $N$, the number of links as $M$, and the minimal hop distance between the server and clients as $H$. This table shows the significant improvements ranging from $17\%$ to $56\%$.

Next, we control three factors which characterize the network topology and reflects the dynamics of the network, to examine their influence on the performance gain of dynamic flow scheduling. Our observation and analysis are as follows.

\begin{table}[t]\footnotesize
  \caption{Factors of network topology that affect performance gain in flow schedule.}
  \label{tab:topology}
  \centering
  \begin{threeparttable}
  \begin{tabular}{ccccc}
    \toprule
    Dataset		&	$N$		&	$M$		& $H$ 	& Performance Gain (\%) \\
    \midrule
   		Waxman		&	1000&	2000	&	8	& 136.6 \\
   		TS\tnote{1}			&3000	&	6020	&	17	& 117.30 \\
	 	Real 1		&	22963	&	48436	&	8	& 131.96 \\
	 	Real 2		&	190920	&	607616	&	15	& 128.12 \\
   		Real 3  	&	54		&	58		&	6	& 156.55 \\
    		Real 4		&	135		&	338		&	4	& 118.47 \\
    \bottomrule
  \end{tabular}
  	\begin{tablenotes}
        \footnotesize
        \item[1] We choose 10 ASes, each AS with 300 nodes. 
      \end{tablenotes}
  \end{threeparttable}
  \vspace{-4mm}
\end{table}

\begin{figure*}[htbp] 
  \begin{minipage}[t]{0.32\linewidth}
    \centering
    \includegraphics [width=\linewidth]{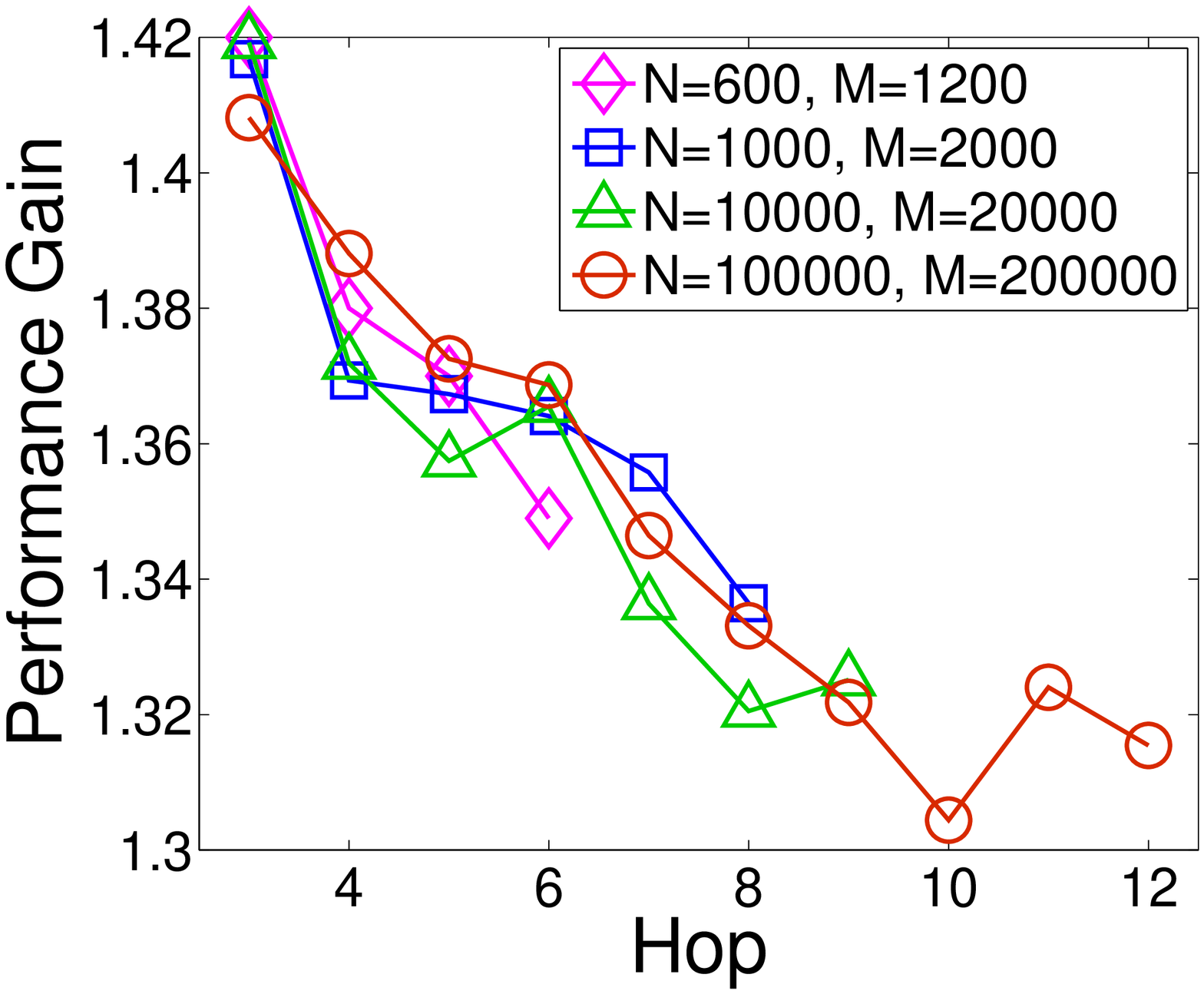}
    \centerline{\scriptsize (a) Impact of topology scale}
  \end{minipage}
  \hfill
  \begin{minipage}[t]{0.32\linewidth}
    \centering
    \includegraphics [width=\linewidth]{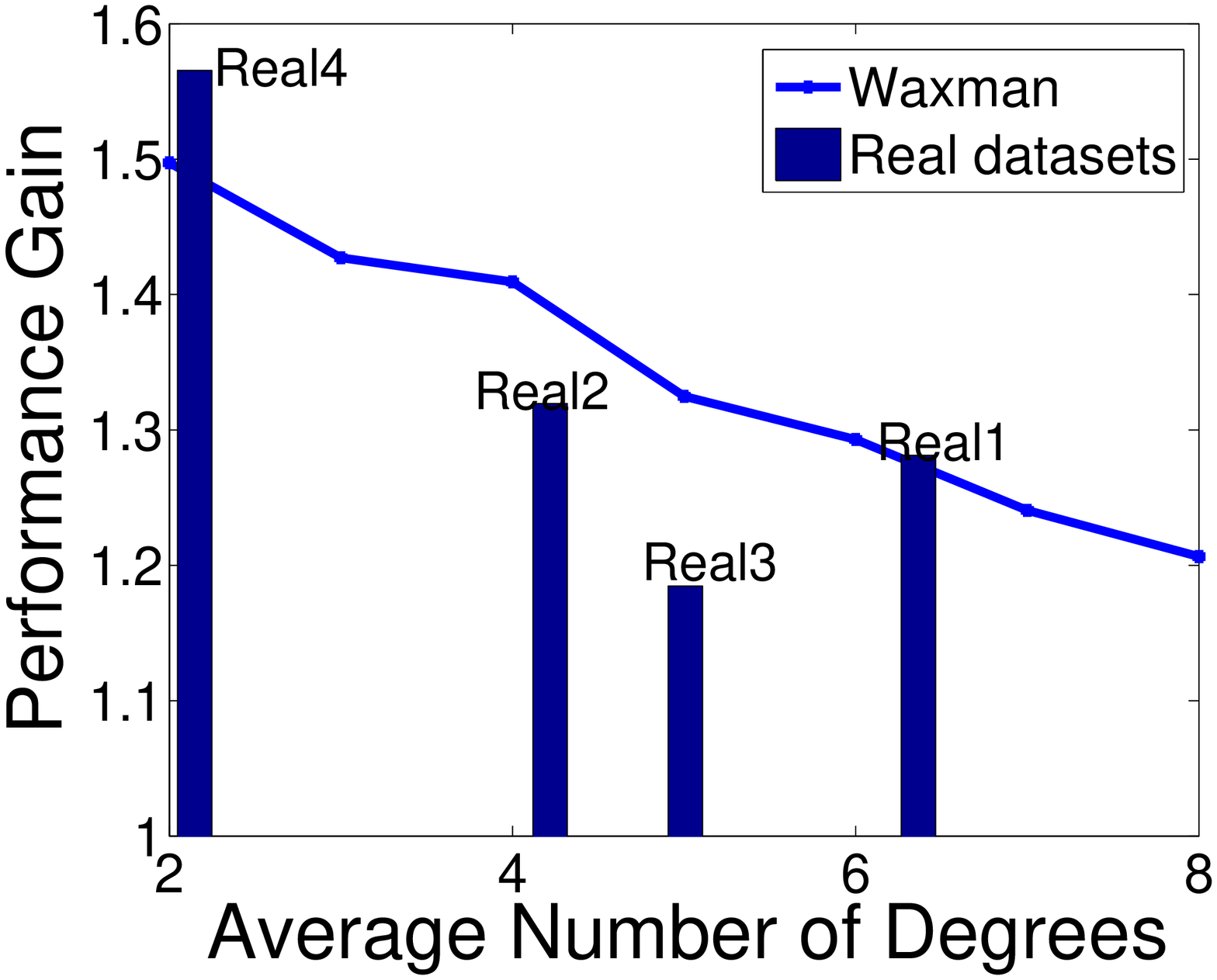}
    \centerline{\scriptsize (b) Impact of node degrees}
  \end{minipage}
  \hfill
    \begin{minipage}[t]{0.32\linewidth}
    \centering
      \includegraphics [width=\linewidth]{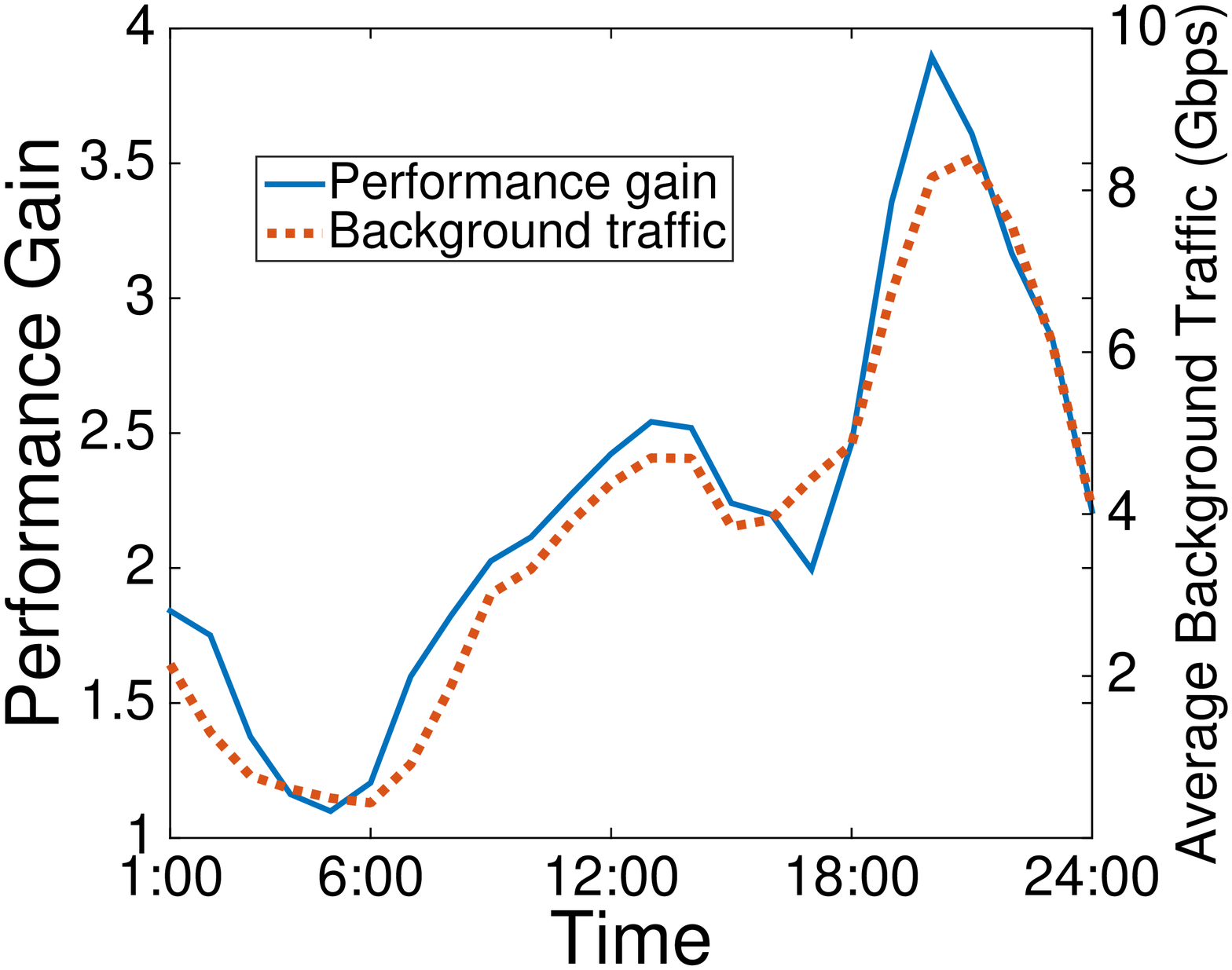}
    \centerline{\scriptsize (c) Impact of network dynamics}
  \end{minipage}
  \hfill
   \vspace{-1mm}
  \caption{Factors of network topology structure and network dynamics that affect performance gain in multihoming flow scheduling.}
  \label{fig:factors1}
  \vspace{-4mm}
\end{figure*}

\textbf{More hops between server and clients lead to lower performance gain.} BRITE allows us to generate topologies with different scales, i.e., different number of nodes and links between them. When the topology scales varying from $600$ nodes to $100,000$ nodes, the number of hops between a server and client networks is varying from $3$ to $12$. In Fig.~\ref{fig:factors1}(a), we observe a general trend that a larger number of hops between clients and a server leads to a smaller performance gain, i.e., the performance gain drops by around $10\%$ when the number of hops increases from $3$ to $10$.

\textbf{Larger average node degree leads to lower performance gain.} In Fig.~\ref{fig:factors1}(b), we plot the performance gain against the average node degrees, using topologies both generated and from real-world. We observe that the performance gain is usually smaller in topologies with a larger number of degrees. For example, the performance gain decreases by $20\%$ when the number of average degrees increases from $2$ to $7$.

\textbf{Higher background load in the network leads to higher performance gain.} 
 We set the average background traffic in each hour based on the realword network workload as showed in Fig.~\ref{fig:factors1}(c)}. When the average background traffic increases, the performance gain improves, which indicates that the dynamic flow schedule can potentially make full use of the available capacity in a network topology.

\section{Dynamical Flow Scheduling in Multihoming Video CDN} \label{sec:design}

In this section, we design a local selection scheme to schedule flows from a multihoming CDN server to client networks. Firstly, we using an information gain approach to discover a set of parameters that have the most important impact on server throughput. Secondly, we present our local heuristic strategy, i.e, form a parameter combination to choose the outgoing ISP, which maximizes the information gain, so as to eventually improve the server throughput.

Next, we present the details of our design.

\subsection{Parameter Selection Based on Measurement Insights}

Based on our previous experiments and analysis in Sec.~\ref{sec:topologyanalysis}, we study the path characteristics between the server and clients, and design the following parameters for the flow scheduling. We define $TOPO_{k}$ as the set of links and nodes (e.g., routers) in the network topology that are involved in a flow scheduling strategy $k$. The parameters below are calculated from $TOPO_{k}$.

$\rhd$ \textbf{Parameters of Network Topology.} The parameters of a network topology are as follows, which can be collected by the server using network measurement tools like Traceroute: 1) $P_{k}$: the sum of links in all routing paths from the server to clients; 2) $E_{k}$: the number of unique links in all routing paths from the server to clients; 3) $O_{k}=(P_{k}-E_{k})$, which represents an ``overlap'' level of links among the paths to all the clients.


$\rhd$ \textbf{Parameters of Network Dynamics.} The parameters of network dynamics are designed as follows, which can be collected using tools like Pathneck \cite{Hu:2004:LIB:1015467.1015474}: 1) $B_{k}$: the number of ``bottlenecks'' in the delivery paths. A bottleneck is a link whose capacity constraints the video traffic delivery; 2) $W_{k}$: the sum of bandwidths of all bottleneck links. 
Furthermore, We split the paths between peering server and clients into three parts, and define parameter: 
3) $BL1_{k}$: the number of bottlenecks that are located nearby the peering server;	
4) $BL2_{k}$: the number of bottlenecks that are located in network;	
5) $BL3_{k}$: the number of bottlenecks that are located nearby the clients.

To illustrate these parameters, Fig.~\ref{fig:4-2} represents the strategy of $k=(A,C,C)$. In this topology, the solid segments represent the links that are actually used for delivering the flows (e.g., the flow goes through a path $A \rightarrow 1 \rightarrow 2 \rightarrow 4$ to client 1), and the slashes are bottlenecks. Then we calculate the value of these parameters as follows: $P_k = 9$, $E_k=7$, $O_k=2$, $B_k=2$, $W_k=6+4=10$, $BL1_{k}=1$, $BL2_{k}=0$, and $BL3_{k}=1$.

\vspace{3mm}
 \makeatletter\def\@captype{figure}\makeatother 
\begin{minipage}{0.4\textwidth} 
\centering 
\includegraphics[width=0.8\textwidth,height=3.5cm]{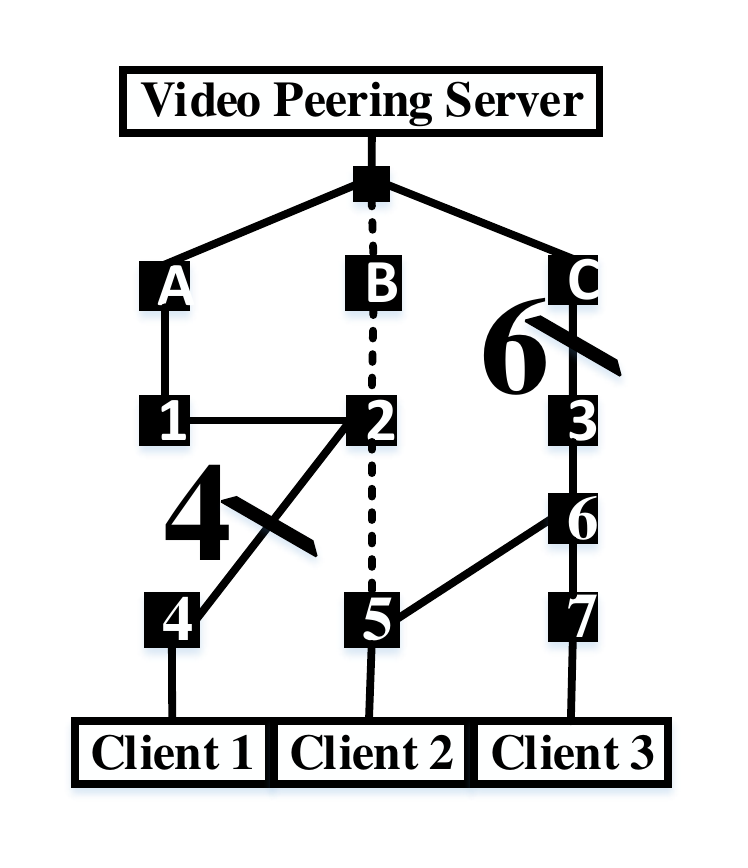} 
\vbox{
 	
 	\tiny $P_k=9$, $E_k=7$\\
 	\tiny $O_k=2$, $B_k=2$\\
 	\tiny $W_k=10$, $BL1_k=1$\\
 	\tiny $BL2_k=0$, $BL3_k=1$
 	}
\caption{Strategy $k=(A,C,C)$.}
 \label{fig:4-2}
\end{minipage} 
\makeatletter\def\@captype{table}\makeatother 
\begin{minipage}{.5\textwidth} 
\footnotesize
\centering 
\caption{Mutual information gain between throughput and parameters.}
 \label{tab:information-gain}
 
  \begin{tabular}{cccccc}
    \toprule
   Metric					&	Waxman	&  TS 		&	Real 1			\\
    \midrule
 		$E_{k}$				&	\textbf{0.305}	&	\textbf{0.653}	&\textbf{0.289}	\\
 		$P_{k}$  			&	0.195			&	0.541			& 0.254	\\ 		
 		$O_{k}$				&	\textbf{0.337}	&	\textbf{0.612}	&\textbf{0.340}	\\
 		$B_{k}$				&	0.228			&	0.412			& 0.296	\\
		$W_{k}$				&	\textbf{0.753}	&	\textbf{0.986}	&\textbf{0.945}	\\ 		
 		$BL1$				&	0.160			&	0.234			&0.264	\\
 		$BL2$				&	0.110			&	0.298			&0.129	\\
 		$BL3$				&	0.098			&	0.184			&0.165	\\
    \bottomrule
  \end{tabular}
\end{minipage} 

\vspace{6mm}
 
  We use an information gain approach \cite{Balachandran:2013:DPM:2486001.2486025} to identify the correlation between the parameters designed above and the throughput. The information gain of the parameter and the throughput is calculated as $I(\beta) = \frac{H(Y)-H(Y|\beta)}{H(Y)}$, where $X$ denotes a parameter which we design, $Y$ is fixed to be the throughput, and $H(X)$ denotes the entropy of the variable $X$. Tab.~\ref{tab:information-gain} shows the \emph{information gain} of the parameters and the throughputs with the dataset of Waxman, TS and Real $1$. We observe that parameters $E_{k}$, $O_{k}$ and $W_{k}$ tend to have the most significant impact on the throughput. To this end, we use them in our flow scheduling.

\subsection{Local Heuristic Strategy for Flow Scheduling}

Based on the selected parameters, $E_{k}$, $O_{k}$ and $W_{k}$, we perform local selection scheme to schedule flows with a combination of parameters. A parameter combination $\mathcal{X} \in \mathbf{P}(\{E_k, O_k, W_k\})$, where $\mathbf{P}(\cdot)$ is a power set function. For a given parameter combination $\mathcal{X}$, a strategy $k$ satisfies the combination if $k$ optimize all the parameters. Since parameters $E_{k}$, $W_{k}$ have positive correlations with the throughput, $O_{k}$ has negative correlation with the throughput, the optimization goals for these parameters are different. For example, if the parameter combination is $\{E_k, O_k\}$, a strategy $k$, which maximizes the number of links in the delivery paths and minimizes the number of overlapped links, is a valid strategy.

Next section, we run experiments to demonstrate the effectiveness of the different combinations of parameters.

\section{Evaluation} \label{sec:evaluation}

In this section, we verify the effectiveness of our design, and run experiments to demonstrate the performance of different combinations of parameters. We let param performance gain denote the performance gain achieved by strategies that satisfy the parameter combinations.

$\rhd$Effectiveness of Parameter Combination: We present the param performance gain achieved by strategies that satisfy the parameter combinations in Tab.~\ref{tab:Flow scheduling gain by using different factors},  The ``baseline'' is the maximum performance gain. We observe that such simple parameters can significantly help a multihoming CDN to schedule flows efficiently. In particular, the param performance gain with a set of parameters is usually better than a single parameter. For example, in the Real 1 topology dataset, using a single parameter $E_{k}$ to guide the flow scheduling achieves up to $40\%$ of the optimal performance gain, while a combination of $\{E_{k},O_{k},W_{k}\}$ can achieve $90\%$ of the optimal improvement.


\vspace{3mm}
\makeatletter\def\@captype{table}\makeatother 
\begin{minipage}{.5\textwidth} 
\scriptsize
\caption{Effectiveness of the selected parameters: param performance gain.}
 \label{tab:Flow scheduling gain by using different factors}
  \begin{tabular}{cccc}
    \toprule
    Metric				&	Waxman(\%)	 &  TS(\%) &Real 1(\%)\\
    \midrule
      Baseline		&	136.6	&117.30		&131.96		\\ \hline
      $\{E_{k}\}$			&	112.80	&107.29		&112.49	\\ \hline
      $\{O_{k}\}$			&	127.60	&111.51		&118.59\\ \hline
      $\{W_{k}\}$			&	115.95	&112.63		&125.87	\\ \hline
      $\{E_{k},O_{k}\} $	&	128.96	&113.23		&123.00	\\  \hline
      $\{E_{k},W_{k}\}$	&   124.15	&113.82		&127.90 \\ \hline
      $\{O_{k},W_{k}\}$&	130.33	&113.43		&124.79	\\ \hline
      $\{E_{k},O_{k},W_{k}\}$ &	130.63	&115.06	& 128.97	\\
    \bottomrule
  \end{tabular}
\end{minipage} 
\makeatletter\def\@captype{figure}\makeatother 
\begin{minipage}{0.4\textwidth} 
\centering 
\includegraphics[width=\textwidth]{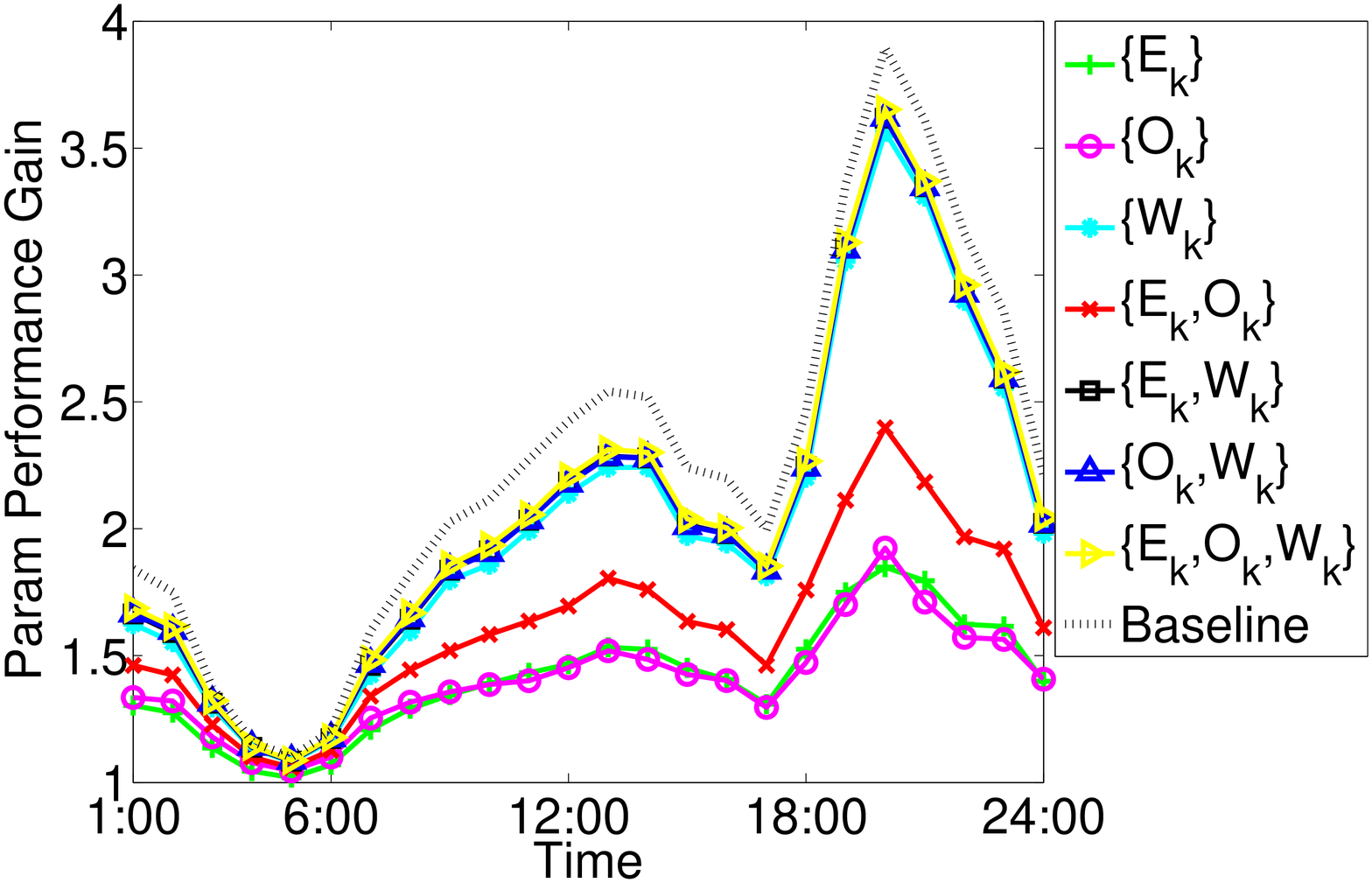} 
\caption{Effectiveness of the selected parameters: param performance gain.}
\label{fig:dynamic Evaluation}
\end{minipage} 
\vspace{6mm}

$\rhd$ Performance under Dynamic Networks: We study the param performance gain in a dynamic network condition (i.e., changing background traffic over links). In Fig.~\ref{fig:dynamic Evaluation}, we plot the param performance gain with a varying background traffic over time. We observe that the performance gain is up to $32\%$-$98\%$ of the optimal performance gain (baseline) at $8$ pm. We also observe that parameter $W_{k}$ plays an important role when the background traffic level is high. When the background traffic level is low (e.g., around 6 am), all these parameters achieve relatively similar performance gain.
  \vspace{-3mm}
\begin{figure}[htbp]

		\begin{minipage}[t]{0.45\linewidth}
    \centering
    \includegraphics [width=\linewidth,height=4cm]{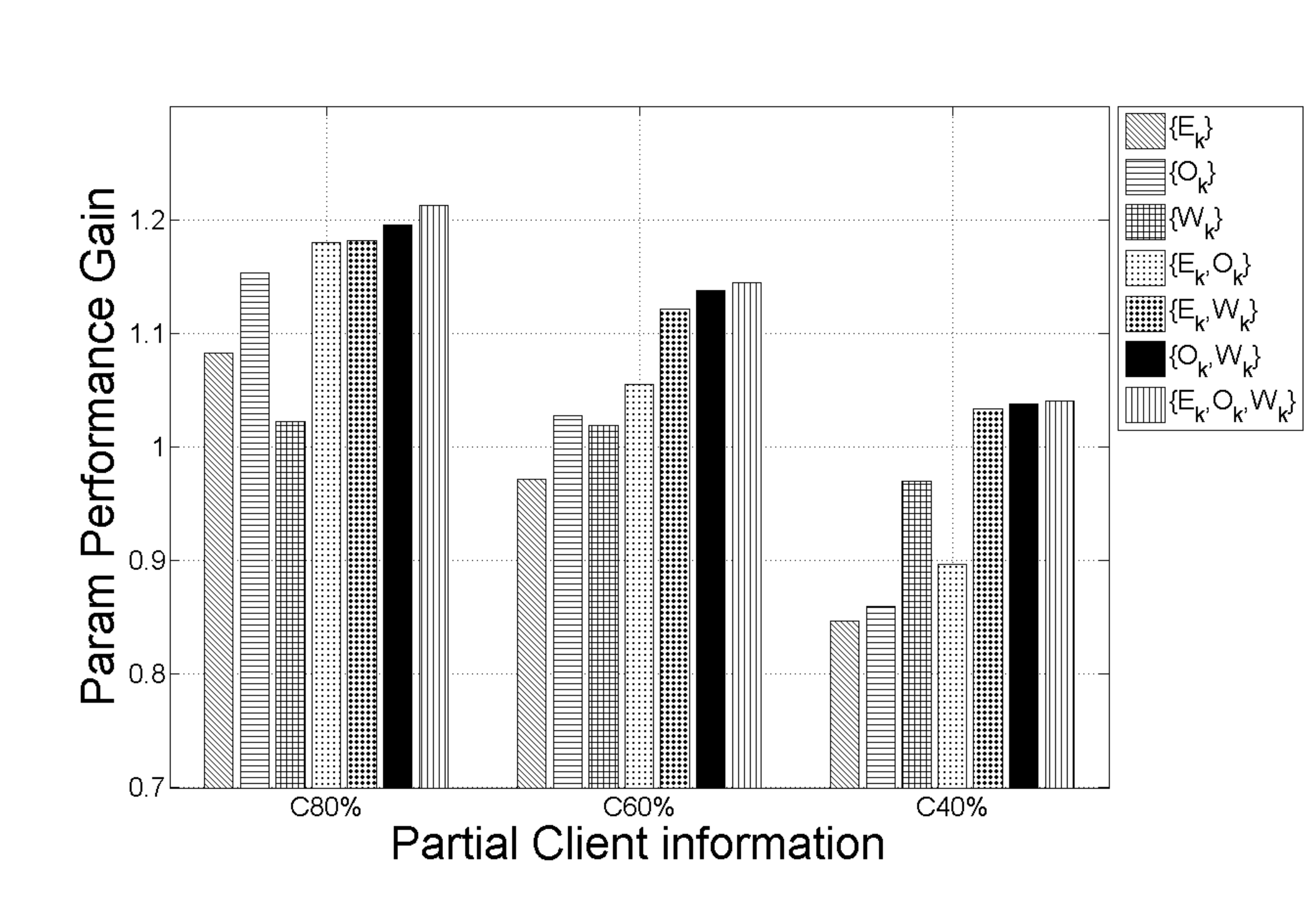}
    \centerline{\scriptsize (a) Information loss of client networks.}
  	\end{minipage}
 	\hfill
 		\begin{minipage}[t]{0.45\linewidth}
    \centering
    \includegraphics [width=\linewidth,height=4cm]{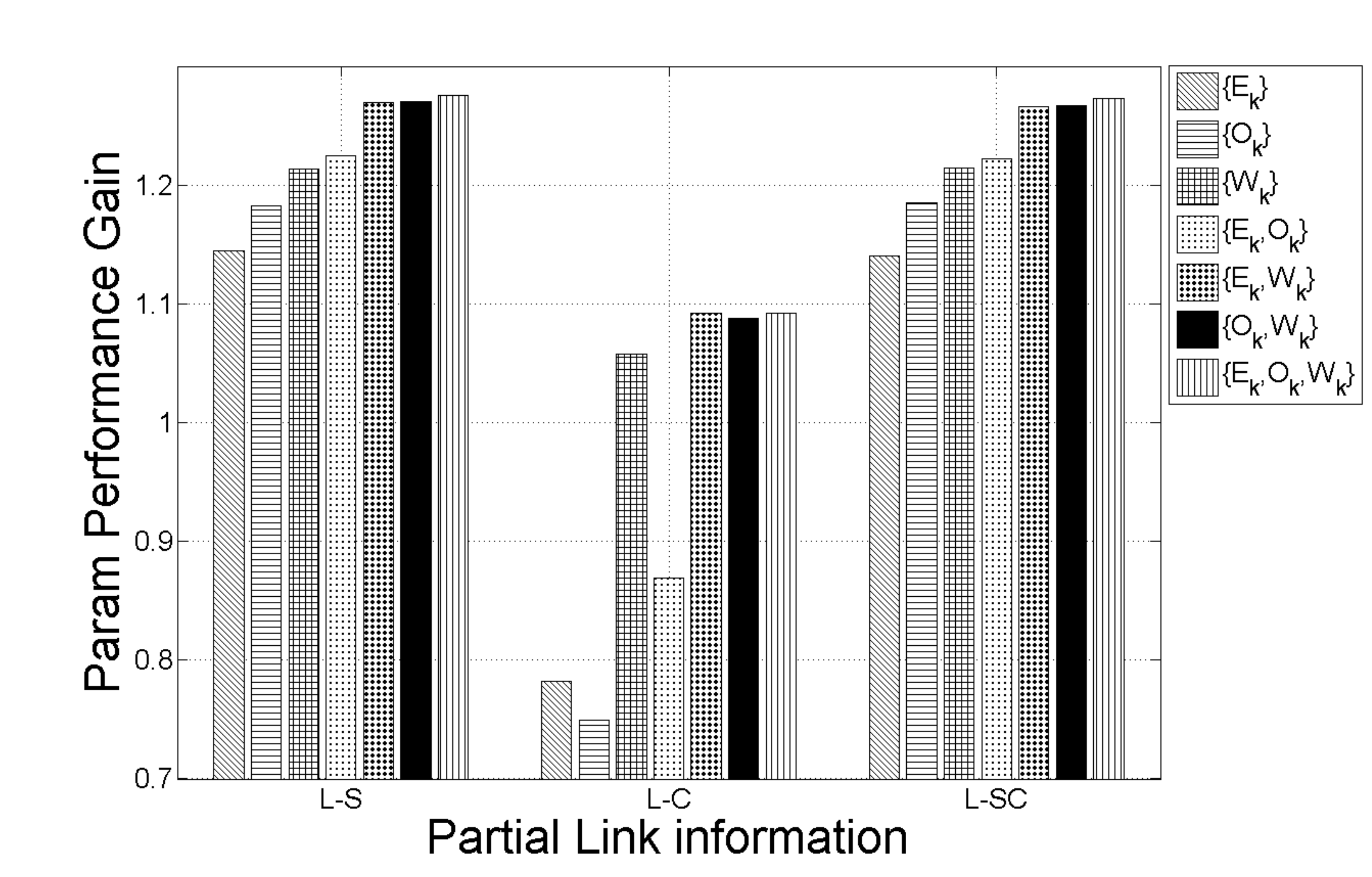}
    \centerline{\scriptsize (b) Information loss of links.}
  	\end{minipage}
 	\hfill
 	
 	\vspace{-1mm}
 	 \caption{Impact of Incomplete Information.}
  \label{fig:dynamic Evaluation1}
  \vspace{-6mm}
\end{figure}

\subsection{Impact of Incomplete Information}
\label{sec:Partial information}
 
In a real-world deployment, it may not be easy to collect complete information on all the parameters. We verify the effectiveness of the local selection scheme when information is incomplete.

$\rhd$ \textbf{Partial client information.} We first evaluate the performance when some client information is missing. Fig.~\ref{fig:dynamic Evaluation1}(a) shows the param performance gain when we calculate the parameters based on $80\%$, $60\%$ and $40\%$ of the client' information. We observe that performance gain decreases as the information loss increases. In Particular, there is almost no performance gain anymore when the client information loss is $60\%$.

$\rhd$ \textbf{Partial link information.} We may also lose information on links. We split the paths between the server and clients into three parts. \emph{L-S} indicates we only have information about $2/3$ links that are close to the server, \emph{L-C} indicates we only have information about $2/3$ links that are close to the clients, and \emph{L-SC} indicates that we have information about $1/3$ links close to the server and $1/3$ links close to the clients. From Fig.~\ref{fig:dynamic Evaluation1}(b), we observe that the information of links close to the server is more important, e.g., \emph{L-C} misses the link information nearby the peering server, and its performance is obviously lower than the performance of the other two. 

\section{Conclusion} \label{sec:conclusion}

Multihoming has become a norm for video CDNs to improve the performance of content delivery. Our real-world experiments reveal that conventional rule-based flow scheduling strategies fail to provide good throughput in dynamic network conditions. In this paper, we study the impact of different network factors, including the network topology and the network dynamics, on the performance in dynamic flow scheduling, in a multihoming CDN. Based on our experiments studies, a set of parameters are selected to guide flow schedule in multihoming CDNs. We use an information gain approach to identify the importance of different parameters, and form a parameter combination to generate the flow scheduling strategy. Our evaluation results show the effectiveness of parameter combination, which achieves a throughput improvement up to 90\% of the optimal solution.

{

\footnotesize
\addtolength{\itemsep}{-6ex}
\bibliographystyle{splncs03}
\bibliography{mymsdn}
}

\end{document}